\catcode`\@=11
\font\tensmc=cmcsc10      
\def\smc{\tensmc}

\def\hcorrection#1{\advance\hoffset by #1 }
\def\vcorrection#1{\advance\voffset by #1 }
\def\wlog#1{}
\newif\iftitle@
\outer\def\title{\title@true\vglue 24\p@ plus 12\p@ minus 12\p@
   \bgroup\let\\=\cr\tabskip\centering
   \halign to \hsize\bgroup\tenbf\hfill\ignorespaces##\unskip\hfill\cr}
\def\endtitle{\cr\egroup\egroup\vglue 18\p@ plus 12\p@ minus 6\p@}
\outer\def\author{\iftitle@\vglue -18\p@ plus -12\p@ minus -6\p@\fi\vglue
    12\p@ plus 6\p@ minus 3\p@\bgroup\let\\=\cr\tabskip\centering
    \halign to \hsize\bgroup\smc\hfill\ignorespaces##\unskip\hfill\cr}
\def\endauthor{\cr\egroup\egroup\vglue 18\p@ plus 12\p@ minus 6\p@}
\outer\def\heading{\bigbreak\bgroup\let\\=\cr\tabskip\centering
    \halign to \hsize\bgroup\smc\hfill\ignorespaces##\unskip\hfill\cr}
\def\endheading{\cr\egroup\egroup\nobreak\medskip}
\outer\def\subheading#1{\medbreak\noindent{\tenbf\ignorespaces
      #1\unskip.\enspace}\ignorespaces}

\outer\def\endproclaim{\par\ifdim\lastskip<\medskipamount\removelastskip
  \penalty 55 \fi\medskip\rm}
\outer\def\demo#1{\par\ifdim\lastskip<\smallskipamount\removelastskip
    \smallskip\fi\noindent{\smc\ignorespaces#1\unskip:\enspace}\rm
      \ignorespaces}

\newcount\footmarkcount@
\footmarkcount@=1
\def\makefootnote@#1#2{\insert\footins{\interlinepenalty=100
  \splittopskip=\ht\strutbox \splitmaxdepth=\dp\strutbox 
  \floatingpenalty=\@MM
  \leftskip=\z@\rightskip=\z@\spaceskip=\z@\xspaceskip=\z@
  \noindent{#1}\footstrut\rm\ignorespaces #2\strut}}
\def\footnote{\let\@sf=\empty\ifhmode\edef\@sf{\spacefactor
   =\the\spacefactor}\/\fi\futurelet\next\footnote@}
\def\footnote@{\ifx"\next\let\next\footnote@@\else
    \let\next\footnote@@@\fi\next}
\def\footnote@@"#1"#2{#1\@sf\relax\makefootnote@{#1}{#2}}
\def\footnote@@@#1{$^{\number\footmarkcount@}$\makefootnote@
   {$^{\number\footmarkcount@}$}{#1}\global\advance\footmarkcount@ by 1 }

\hyphenation{man-u-script man-u-scripts ap-pen-dix ap-pen-di-ces}
\hyphenation{data-base data-bases}
\ifx\amstexloaded@\relax\catcode`\@=13 
  \endinput\else\let\amstexloaded@=\relax\fi
\newlinechar=`\^^J
\def\eat@#1{}
\def\Space@.{\futurelet\Space@\relax}
\Space@. %
\newhelp\athelp@
{Only certain combinations beginning with @ make sense to me.^^J
Perhaps you wanted \string\@\space for a printed @?^^J
I've ignored the character or group after @.}
\def\futureletnextat@{\futurelet\next\at@}
{\catcode`\@=\active
\lccode`\Z=`\@ \lowercase
{\gdef@{\expandafter\csname futureletnextatZ\endcsname}
\expandafter\gdef\csname atZ\endcsname
   {\ifcat\noexpand\next a\def\next{\csname atZZ\endcsname}\else
   \ifcat\noexpand\next0\def\next{\csname atZZ\endcsname}\else
    \def\next{\csname atZZZ\endcsname}\fi\fi\next}
\expandafter\gdef\csname atZZ\endcsname#1{\expandafter
   \ifx\csname #1Zat\endcsname\relax\def\next
     {\errhelp\expandafter=\csname athelpZ\endcsname
      \errmessage{Invalid use of \string@}}\else
       \def\next{\csname #1Zat\endcsname}\fi\next}
\expandafter\gdef\csname atZZZ\endcsname#1{\errhelp
    \expandafter=\csname athelpZ\endcsname
      \errmessage{Invalid use of \string@}}}}
\def\atdef@#1{\expandafter\def\csname #1@at\endcsname}
\newhelp\defahelp@{If you typed \string\define\space cs instead of
\string\define\string\cs\space^^J
I've substituted an inaccessible control sequence so that your^^J
definition will be completed without mixing me up too badly.^^J
If you typed \string\define{\string\cs} the inaccessible control sequence^^J
was defined to be \string\cs, and the rest of your^^J
definition appears as input.}
\newhelp\defbhelp@{I've ignored your definition, because it might^^J
conflict with other uses that are important to me.}
\def\define{\futurelet\next\define@}
\def\define@{\ifcat\noexpand\next\relax
  \def\next{\define@@}%
  \else\errhelp=\defahelp@
  \errmessage{\string\define\space must be followed by a control 
     sequence}\def\next{\def\garbage@}\fi\next}
\def\undefined@{}
\def\preloaded@{}    
\def\define@@#1{\ifx#1\relax\errhelp=\defbhelp@
   \errmessage{\string#1\space is already defined}\def\next{\def\garbage@}%
   \else\expandafter\ifx\csname\expandafter\eat@\string
         #1@\endcsname\undefined@\errhelp=\defbhelp@
   \errmessage{\string#1\space can't be defined}\def\next{\def\garbage@}%
   \else\expandafter\ifx\csname\expandafter\eat@\string#1\endcsname\relax
     \def\next{\def#1}\else\errhelp=\defbhelp@
     \errmessage{\string#1\space is already defined}\def\next{\def\garbage@}%
      \fi\fi\fi\next}
\def\famzero{\fam\z@}

\def\exp{\mathop{\famzero exp}\nolimits}

\def\lim{\mathop{\famzero lim}}

\def\textfont@#1#2{\def#1{\relax\ifmmode
    \errmessage{Use \string#1\space only in text}\else#2\fi}}
\textfont@\rm\tenrm
\textfont@\it\tenit
\textfont@\sl\tensl
\textfont@\bf\tenbf
\textfont@\smc\tensmc
\let\ic@=\/
\def\/{\unskip\ic@}
\def\textfonti{\the\textfont1 }
\def\t#1#2{{\edef\next{\the\font}\textfonti\accent"7F \next#1#2}}
\let\B=\=
\let\D=\.
\def~{\unskip\nobreak\ \ignorespaces}
{\catcode`\@=\active
\gdef\@{\char'100 }}
\atdef@-{\leavevmode\futurelet\next\athyph@}
\def\athyph@{\ifx\next-\let\next=\athyph@@
  \else\let\next=\athyph@@@\fi\next}
\def\athyph@@@{\hbox{-}}
\def\athyph@@#1{\futurelet\next\athyph@@@@}
\def\athyph@@@@{\if\next-\def\next##1{\hbox{---}}\else
    \def\next{\hbox{--}}\fi\next}
\def\.{.\spacefactor=\@m}
\atdef@.{\null.}
\atdef@,{\null,}
\atdef@;{\null;}
\atdef@:{\null:}
\atdef@?{\null?}
\atdef@!{\null!}   
\def\srdr@{\thinspace}                     
\def\drsr@{\kern.02778em}
\def\sldl@{\kern.02778em}
\def\dlsl@{\thinspace}
\atdef@"{\unskip\futurelet\next\atqq@}
\def\atqq@{\ifx\next\Space@\def\next. {\atqq@@}\else
         \def\next.{\atqq@@}\fi\next.}
\def\atqq@@{\futurelet\next\atqq@@@}
\def\atqq@@@{\ifx\next`\def\next`{\atqql@}\else\def\next'{\atqqr@}\fi\next}
\def\atqql@{\futurelet\next\atqql@@}
\def\atqql@@{\ifx\next`\def\next`{\sldl@``}\else\def\next{\dlsl@`}\fi\next}
\def\atqqr@{\futurelet\next\atqqr@@}
\def\atqqr@@{\ifx\next'\def\next'{\srdr@''}\else\def\next{\drsr@'}\fi\next}

\def\textfontii{\the\textfont2 }
\def\{{\relax\ifmmode\lbrace\else
    {\textfontii f}\spacefactor=\@m\fi}
\def\}{\relax\ifmmode\rbrace\else
    \let\@sf=\empty\ifhmode\edef\@sf{\spacefactor=\the\spacefactor}\fi
      {\textfontii g}\@sf\relax\fi}   
\def\nonhmodeerr@#1{\errmessage
     {\string#1\space allowed only within text}}
\def\linebreak{\relax\ifhmode\unskip\break\else
    \nonhmodeerr@\linebreak\fi}
\def\allowlinebreak{\relax
   \ifhmode\allowbreak\else\nonhmodeerr@\allowlinebreak\fi}
\newskip\saveskip@
\def\nolinebreak{\relax\ifhmode\saveskip@=\lastskip\unskip
  \nobreak\ifdim\saveskip@>\z@\hskip\saveskip@\fi
   \else\nonhmodeerr@\nolinebreak\fi}
\def\newline{\relax\ifhmode\null\hfil\break
    \else\nonhmodeerr@\newline\fi}
\def\nonmathaerr@#1{\errmessage
     {\string#1\space is not allowed in display math mode}}
\def\nonmathberr@#1{\errmessage{\string#1\space is allowed only in math mode}}
\def\mathbreak{\relax\ifmmode\ifinner\break\else
   \nonmathaerr@\mathbreak\fi\else\nonmathberr@\mathbreak\fi}
\def\nomathbreak{\relax\ifmmode\ifinner\nobreak\else
    \nonmathaerr@\nomathbreak\fi\else\nonmathberr@\nomathbreak\fi}
\def\allowmathbreak{\relax\ifmmode\ifinner\allowbreak\else
     \nonmathaerr@\allowmathbreak\fi\else\nonmathberr@\allowmathbreak\fi}
\def\pagebreak{\relax\ifmmode
   \ifinner\errmessage{\string\pagebreak\space
     not allowed in non-display math mode}\else\postdisplaypenalty-\@M\fi
   \else\ifvmode\penalty-\@M\else\edef\spacefactor@
       {\spacefactor=\the\spacefactor}\vadjust{\penalty-\@M}\spacefactor@
        \relax\fi\fi}
\def\nopagebreak{\relax\ifmmode
     \ifinner\errmessage{\string\nopagebreak\space
    not allowed in non-display math mode}\else\postdisplaypenalty\@M\fi
    \else\ifvmode\nobreak\else\edef\spacefactor@
        {\spacefactor=\the\spacefactor}\vadjust{\penalty\@M}\spacefactor@
         \relax\fi\fi}
\def\newpage{\relax\ifvmode\vfill\penalty-\@M\else\nonvmodeerr@\newpage\fi}
\def\nonvmodeerr@#1{\errmessage
    {\string#1\space is allowed only between paragraphs}}
\def\smallpagebreak{\relax\ifvmode\smallbreak
      \else\nonvmodeerr@\smallpagebreak\fi}
\def\medpagebreak{\relax\ifvmode\medbreak
       \else\nonvmodeerr@\medpagebreak\fi}
\def\bigpagebreak{\relax\ifvmode\bigbreak
      \else\nonvmodeerr@\bigpagebreak\fi}
\newdimen\captionwidth@
\captionwidth@=\hsize
\advance\captionwidth@ by -1.5in
\def\caption#1{}
\def\topspace#1{\gdef\thespace@{#1}\ifvmode\def\next
    {\futurelet\next\topspace@}\else\def\next{\nonvmodeerr@\topspace}\fi\next}
\def\topspace@{\ifx\next\Space@\def\next. {\futurelet\next\topspace@@}\else
     \def\next.{\futurelet\next\topspace@@}\fi\next.}
\def\topspace@@{\ifx\next\caption\let\next\topspace@@@\else
    \let\next\topspace@@@@\fi\next}
 \def\topspace@@@@{\topinsert\vbox to 
       \thespace@{}\endinsert}
\def\topspace@@@\caption#1{\topinsert\vbox to
    \thespace@{}\nobreak
      \smallskip
    \setbox\z@=\hbox{\noindent\ignorespaces#1\unskip}%
   \ifdim\wd\z@>\captionwidth@
   \centerline{\vbox{\hsize=\captionwidth@\noindent\ignorespaces#1\unskip}}%
   \else\centerline{\box\z@}\fi\endinsert}
\def\midspace#1{\gdef\thespace@{#1}\ifvmode\def\next
    {\futurelet\next\midspace@}\else\def\next{\nonvmodeerr@\midspace}\fi\next}
\def\midspace@{\ifx\next\Space@\def\next. {\futurelet\next\midspace@@}\else
     \def\next.{\futurelet\next\midspace@@}\fi\next.}
\def\midspace@@{\ifx\next\caption\let\next\midspace@@@\else
    \let\next\midspace@@@@\fi\next}
 \def\midspace@@@@{\midinsert\vbox to 
       \thespace@{}\endinsert}
\def\midspace@@@\caption#1{\midinsert\vbox to
    \thespace@{}\nobreak
      \smallskip
      \setbox\z@=\hbox{\noindent\ignorespaces#1\unskip}%
      \ifdim\wd\z@>\captionwidth@
    \centerline{\vbox{\hsize=\captionwidth@\noindent\ignorespaces#1\unskip}}%
    \else\centerline{\box\z@}\fi\endinsert}
\mathchardef\prime@="0230
\def\prime{{{}\prime@{}}}
\def\prim@s{\prime@\futurelet\next\pr@m@s}

\def\,{\relax\ifmmode\mskip\thinmuskip\else\thinspace\fi}
\def\!{\relax\ifmmode\mskip-\thinmuskip\else\negthinspace\fi}
\def\frac#1#2{{#1\over#2}}

\def\:{\nobreak\hskip.1111em{:}\hskip.3333em plus .0555em\relax}
\def\intic@{\mathchoice{\hskip5\p@}{\hskip4\p@}{\hskip4\p@}{\hskip4\p@}}
\def\negintic@
 {\mathchoice{\hskip-5\p@}{\hskip-4\p@}{\hskip-4\p@}{\hskip-4\p@}}
\def\intkern@{\mathchoice{\!\!\!}{\!\!}{\!\!}{\!\!}}
\def\intdots@{\mathchoice{\cdots}{{\cdotp}\mkern1.5mu
    {\cdotp}\mkern1.5mu{\cdotp}}{{\cdotp}\mkern1mu{\cdotp}\mkern1mu
      {\cdotp}}{{\cdotp}\mkern1mu{\cdotp}\mkern1mu{\cdotp}}}
\newcount\intno@             
\def\iint{\intno@=\tw@\futurelet\next\ints@} 
\def\iiint{\intno@=\thr@@\futurelet\next\ints@}
\def\iiiint{\intno@=4 \futurelet\next\ints@}
\def\idotsint{\intno@=\z@\futurelet\next\ints@}
\def\ints@{\findlimits@\ints@@}
\newif\iflimtoken@
\newif\iflimits@
\def\findlimits@{\limtoken@false\limits@false\ifx\next\limits
 \limtoken@true\limits@true\else\ifx\next\nolimits\limtoken@true\limits@false
    \fi\fi}
\def\multintlimits@{\intop\ifnum\intno@=\z@\intdots@
  \else\intkern@\fi
    \ifnum\intno@>\tw@\intop\intkern@\fi
     \ifnum\intno@>\thr@@\intop\intkern@\fi\intop}
\def\multint@{\int\ifnum\intno@=\z@\intdots@\else\intkern@\fi
   \ifnum\intno@>\tw@\int\intkern@\fi
    \ifnum\intno@>\thr@@\int\intkern@\fi\int}
\def\ints@@{\iflimtoken@\def\ints@@@{\iflimits@
   \negintic@\mathop{\intic@\multintlimits@}\limits\else
    \multint@\nolimits\fi\eat@}\else
     \def\ints@@@{\multint@\nolimits}\fi\ints@@@}
\def\Sb{_\bgroup\vspace@
        \baselineskip=\fontdimen10 \scriptfont\tw@
        \advance\baselineskip by \fontdimen12 \scriptfont\tw@
        \lineskip=\thr@@\fontdimen8 \scriptfont\thr@@
        \lineskiplimit=\thr@@\fontdimen8 \scriptfont\thr@@
        \Let@\vbox\bgroup\halign\bgroup \hfil$\scriptstyle
            {##}$\hfil\cr}
\def\endSb{\crcr\egroup\egroup\egroup}
\def\Sp{^\bgroup\vspace@
        \baselineskip=\fontdimen10 \scriptfont\tw@
        \advance\baselineskip by \fontdimen12 \scriptfont\tw@
        \lineskip=\thr@@\fontdimen8 \scriptfont\thr@@
        \lineskiplimit=\thr@@\fontdimen8 \scriptfont\thr@@
        \Let@\vbox\bgroup\halign\bgroup \hfil$\scriptstyle
            {##}$\hfil\cr}
\def\endSp{\crcr\egroup\egroup\egroup}
\def\Let@{\relax\iffalse{\fi\let\\=\cr\iffalse}\fi}
\def\vspace@{\def\vspace##1{\noalign{\vskip##1 }}}
\def\aligned{\,\vcenter\bgroup\vspace@\Let@\openup\jot\m@th\ialign
  \bgroup \strut\hfil$\displaystyle{##}$&$\displaystyle{{}##}$\hfil\crcr}
\def\endaligned{\crcr\egroup\egroup}
\def\matrix{\,\vcenter\bgroup\Let@\vspace@
    \normalbaselines
  \m@th\ialign\bgroup\hfil$##$\hfil&&\quad\hfil$##$\hfil\crcr
    \mathstrut\crcr\noalign{\kern-\baselineskip}}
\def\endmatrix{\crcr\mathstrut\crcr\noalign{\kern-\baselineskip}\egroup
                \egroup\,}
\newtoks\hashtoks@
\hashtoks@={#}
\def\format{\crcr\egroup\iffalse{\fi\ifnum`}=0 \fi\format@}
\def\format@#1\\{\def\preamble@{#1}%
  \def\c{\hfil$\the\hashtoks@$\hfil}%
  \def\r{\hfil$\the\hashtoks@$}%
  \def\l{$\the\hashtoks@$\hfil}%
  \setbox\z@=\hbox{\xdef\Preamble@{\preamble@}}\ifnum`{=0 \fi\iffalse}\fi
   \ialign\bgroup\span\Preamble@\crcr}

\def\cases{\left\{\,\vcenter\bgroup\vspace@
     \normalbaselines\openup\jot\m@th
       \Let@\ialign\bgroup$##$\hfil&\quad$##$\hfil\crcr
      \mathstrut\crcr\noalign{\kern-\baselineskip}}

\newif\iftagsleft@
\tagsleft@true
\def\TagsOnRight{\global\tagsleft@false}
\def\tag#1$${\iftagsleft@\leqno\else\eqno\fi
 \hbox{\def\pagebreak{\global\postdisplaypenalty-\@M}%
 \def\nopagebreak{\global\postdisplaypenalty\@M}\rm(#1\unskip)}%
  $$\postdisplaypenalty\z@\ignorespaces}
\interdisplaylinepenalty=\@M
\def\allowdisplaybreak@{\def\allowdisplaybreak{\noalign{\allowbreak}}}
\def\displaybreak@{\def\displaybreak{\noalign{\break}}}
\def\align#1\endalign{\def\tag{&}\vspace@\allowdisplaybreak@\displaybreak@
  \iftagsleft@\lalign@#1\endalign\else
   \ralign@#1\endalign\fi}
\def\ralign@#1\endalign{\displ@y\Let@\tabskip\centering\halign to\displaywidth
     {\hfil$\displaystyle{##}$\tabskip=\z@&$\displaystyle{{}##}$\hfil
       \tabskip=\centering&\llap{\hbox{(\rm##\unskip)}}\tabskip\z@\crcr
             #1\crcr}}
\def\lalign@
 #1\endalign{\displ@y\Let@\tabskip\centering\halign to \displaywidth
   {\hfil$\displaystyle{##}$\tabskip=\z@&$\displaystyle{{}##}$\hfil
   \tabskip=\centering&\kern-\displaywidth
        \rlap{\hbox{(\rm##\unskip)}}\tabskip=\displaywidth\crcr
               #1\crcr}}
\def\overrightarrow{\mathpalette\overrightarrow@}
\def\overrightarrow@#1#2{\vbox{\ialign{$##$\cr
    #1{-}\mkern-6mu\cleaders\hbox{$#1\mkern-2mu{-}\mkern-2mu$}\hfill
     \mkern-6mu{\to}\cr
     \noalign{\kern -1\p@\nointerlineskip}
     \hfil#1#2\hfil\cr}}}
\def\overleftarrow{\mathpalette\overleftarrow@}
\def\overleftarrow@#1#2{\vbox{\ialign{$##$\cr
     #1{\leftarrow}\mkern-6mu\cleaders\hbox{$#1\mkern-2mu{-}\mkern-2mu$}\hfill
      \mkern-6mu{-}\cr
     \noalign{\kern -1\p@\nointerlineskip}
     \hfil#1#2\hfil\cr}}}
\def\overleftrightarrow{\mathpalette\overleftrightarrow@}
\def\overleftrightarrow@#1#2{\vbox{\ialign{$##$\cr
     #1{\leftarrow}\mkern-6mu\cleaders\hbox{$#1\mkern-2mu{-}\mkern-2mu$}\hfill
       \mkern-6mu{\to}\cr
    \noalign{\kern -1\p@\nointerlineskip}
      \hfil#1#2\hfil\cr}}}
\def\underrightarrow{\mathpalette\underrightarrow@}
\def\underrightarrow@#1#2{\vtop{\ialign{$##$\cr
    \hfil#1#2\hfil\cr
     \noalign{\kern -1\p@\nointerlineskip}
    #1{-}\mkern-6mu\cleaders\hbox{$#1\mkern-2mu{-}\mkern-2mu$}\hfill
     \mkern-6mu{\to}\cr}}}
\def\underleftarrow{\mathpalette\underleftarrow@}
\def\underleftarrow@#1#2{\vtop{\ialign{$##$\cr
     \hfil#1#2\hfil\cr
     \noalign{\kern -1\p@\nointerlineskip}
     #1{\leftarrow}\mkern-6mu\cleaders\hbox{$#1\mkern-2mu{-}\mkern-2mu$}\hfill
      \mkern-6mu{-}\cr}}}
\def\underleftrightarrow{\mathpalette\underleftrightarrow@}
\def\underleftrightarrow@#1#2{\vtop{\ialign{$##$\cr
      \hfil#1#2\hfil\cr
    \noalign{\kern -1\p@\nointerlineskip}
     #1{\leftarrow}\mkern-6mu\cleaders\hbox{$#1\mkern-2mu{-}\mkern-2mu$}\hfill
       \mkern-6mu{\to}\cr}}}
\def\sqrt#1{\radical"270370 {#1}}
\def\dots{\relax\ifmmode\let\next=\ldots\else\let\next=\tdots@\fi\next}
\def\tdots@{\unskip\ \tdots@@}
\def\tdots@@{\futurelet\next\tdots@@@}
\def\tdots@@@{$\mathinner{\ldotp\ldotp\ldotp}\,
   \ifx\next,$\else
   \ifx\next.\,$\else
   \ifx\next;\,$\else
   \ifx\next:\,$\else
   \ifx\next?\,$\else
   \ifx\next!\,$\else
   $ \fi\fi\fi\fi\fi\fi}
\def\text{\relax\ifmmode\let\next=\text@\else\let\next=\text@@\fi\next}
\def\text@@#1{\hbox{#1}}
\def\text@#1{\mathchoice
 {\hbox{\everymath{\displaystyle}\def\textfonti{\the\textfont1 }%
    \def\textfontii{\the\textfont2 }\textdef@@ T#1}}
 {\hbox{\everymath{\textstyle}\def\textfonti{\the\textfont1 }%
    \def\textfontii{\the\textfont2 }\textdef@@ T#1}}
 {\hbox{\everymath{\scriptstyle}\def\textfonti{\the\scriptfont1 }%
   \def\textfontii{\the\scriptfont2 }\textdef@@ S\rm#1}}
 {\hbox{\everymath{\scriptscriptstyle}\def\textfonti{\the\scriptscriptfont1 }%
   \def\textfontii{\the\scriptscriptfont2 }\textdef@@ s\rm#1}}}
\def\textdef@@#1{\textdef@#1\rm \textdef@#1\bf
   \textdef@#1\sl \textdef@#1\it}

\def\textdef@#1#2{\def\next{\csname\expandafter\eat@\string#2fam\endcsname}%
\if S#1\edef#2{\the\scriptfont\next\relax}%
 \else\if s#1\edef#2{\the\scriptscriptfont\next\relax}%
 \else\edef#2{\the\textfont\next\relax}\fi\fi}
\scriptfont\itfam=\tenit \scriptscriptfont\itfam=\tenit
\scriptfont\slfam=\tensl \scriptscriptfont\slfam=\tensl
\mathcode`\0="0030
\mathcode`\1="0031
\mathcode`\2="0032
\mathcode`\3="0033
\mathcode`\4="0034
\mathcode`\5="0035
\mathcode`\6="0036
\mathcode`\7="0037
\mathcode`\8="0038
\mathcode`\9="0039
\def\Cal{\relax\ifmmode\let\next=\Cal@\else
     \def\next{\errmessage{Use \string\Cal\space only in math mode}}\fi\next}
\def\Cal@#1{{\fam2 #1}}
\def\bold{\relax\ifmmode\let\next=\bold@\else
   \def\next{\errmessage{Use \string\bold\space only in math
      mode}}\fi\next}\def\bold@#1{{\fam\bffam #1}}
\mathchardef\Gamma="0000
\mathchardef\Delta="0001
\mathchardef\Theta="0002
\mathchardef\Lambda="0003
\mathchardef\Xi="0004
\mathchardef\Pi="0005
\mathchardef\Sigma="0006
\mathchardef\Upsilon="0007
\mathchardef\Phi="0008
\mathchardef\Psi="0009
\mathchardef\Omega="000A
\mathchardef\varGamma="0100
\mathchardef\varDelta="0101
\mathchardef\varTheta="0102
\mathchardef\varLambda="0103
\mathchardef\varXi="0104
\mathchardef\varPi="0105
\mathchardef\varSigma="0106
\mathchardef\varUpsilon="0107
\mathchardef\varPhi="0108
\mathchardef\varPsi="0109
\mathchardef\varOmega="010A
\font\dummyft@=dummy
\fontdimen1 \dummyft@=\z@
\fontdimen2 \dummyft@=\z@
\fontdimen3 \dummyft@=\z@
\fontdimen4 \dummyft@=\z@
\fontdimen5 \dummyft@=\z@
\fontdimen6 \dummyft@=\z@
\fontdimen7 \dummyft@=\z@
\fontdimen8 \dummyft@=\z@
\fontdimen9 \dummyft@=\z@
\fontdimen10 \dummyft@=\z@
\fontdimen11 \dummyft@=\z@
\fontdimen12 \dummyft@=\z@
\fontdimen13 \dummyft@=\z@
\fontdimen14 \dummyft@=\z@
\fontdimen15 \dummyft@=\z@
\fontdimen16 \dummyft@=\z@
\fontdimen17 \dummyft@=\z@
\fontdimen18 \dummyft@=\z@
\fontdimen19 \dummyft@=\z@
\fontdimen20 \dummyft@=\z@
\fontdimen21 \dummyft@=\z@
\fontdimen22 \dummyft@=\z@
\def\fontlist@{\\{\tenrm}\\{\sevenrm}\\{\fiverm}\\{\teni}\\{\seveni}%
 \\{\fivei}\\{\tensy}\\{\sevensy}\\{\fivesy}\\{\tenex}\\{\tenbf}\\{\sevenbf}%
 \\{\fivebf}\\{\tensl}\\{\tenit}\\{\tensmc}}
\def\dodummy@{{\def\\##1{\global\let##1=\dummyft@}\fontlist@}}
\newif\ifsyntax@
\newcount\countxviii@
\def\newtoks@{\alloc@5\toks\toksdef\@cclvi}
\def\nopages@{\output={\setbox\z@=\box\@cclv \deadcycles=\z@}\newtoks@\output}
\def\syntax{\syntax@true\dodummy@\countxviii@=\count18
\loop \ifnum\countxviii@ > \z@ \textfont\countxviii@=\dummyft@
   \scriptfont\countxviii@=\dummyft@ \scriptscriptfont\countxviii@=\dummyft@
     \advance\countxviii@ by-\@ne\repeat
\dummyft@\tracinglostchars=\z@
  \nopages@\frenchspacing\hbadness=\@M}
\def\magstep#1{\ifcase#1 1000\or
 1200\or 1440\or 1728\or 2074\or 2488\or 
 \errmessage{\string\magstep\space only works up to 5}\fi\relax}
{\lccode`\2=`\p \lccode`\3=`\t 
 \lowercase{\gdef\tru@#123{#1truept}}}

\def\scaletype#1{\mag=#1\relax
 \hsize=\expandafter\tru@\the\hsize
 \vsize=\expandafter\tru@\the\vsize
 \dimen\footins=\expandafter\tru@\the\dimen\footins}

\def\scalefont#1#2\andcallit#3{\edef\font@{\the\font}#1\font#3=
  \fontname\font\space scaled #2\relax\font@}
\def\Mag@#1#2{\ifdim#1<1pt\multiply#1 #2\relax\divide#1 1000 \else
  \ifdim#1<10pt\divide#1 10 \multiply#1 #2\relax\divide#1 100\else
  \divide#1 100 \multiply#1 #2\relax\divide#1 10 \fi\fi}
\def\scalelinespacing#1{\Mag@\baselineskip{#1}\Mag@\lineskip{#1}%
  \Mag@\lineskiplimit{#1}}
\def\wlog#1{\immediate\write-1{#1}}
\catcode`\@=\active

\magnification 1200
\hsize=15.2truecm
\overfullrule=0pt
\vsize=22.6truecm 

\font\sixrm=cmr6        
\font\eightrm=cmr8

\font\eightsl=cmsl8
\font\gross=cmcsc10

\font\tit=cmti10 scaled\magstep 2      
 2

\hyphenation{
Meas-ure-ment Meas-uring 
meas-ure-ment meas-ure meas-uring
pre-meas-ure-ment pre-meas-ure pre-meas-uring}

\def\sy{system}                       
\def\qu{quantum}                      \def\mec{mechanic}
\def\qm{quantum mechanics}
\def\ty{theory}                       
\def\me{meas\-ure}                    \def\mt{\me{}ment} \def\mts{\mt{}s}

\def\op{operator}                    
\def\ob{observable}                  \def\obs{observables}
\def\sad{self-adjoint}               \def\sop{\sad\ \op}

\def\pos{position}                    \def\mom{momentum}

\def\pee#1{{P[#1]}}           
\def\bo#1{{\bold #1}}                    

\def\ip#1#2{\left\langle\,#1\,|\,#2\,\right\rangle} 
         \def\tst#1{{\textstyle{#1}}}  
    \def\ts12{{\textstyle{\frac 12}}}

\def\Rea{{\bold R}}

\def\fii{\varphi}   
                       \def\cs{$\Cal S$}

\def\hi{{\Cal H}}

\def\pa{\Cal R_{\Cal S}}         \def\pas{\Cal R_{\Cal A}} 

\def\s{$\Cal S$}         \def\a{$\Cal A$}      






\def\us{U(\varphi\otimes\phi )}

\def\vas{\langle}                    \def\oik{\rangle}

\def\var{\text{Var}}                 \def\12pi{\frac 1{2\pi}}
\def\fqf{E^{e}}                    \def\fpg{F^{f}}

\font\eightsl=cmsl8
\font\tit=cmti10 scaled\magstep 1


\advance\baselineskip by 6pt
\advance\hoffset by -1cm
\advance \vsize by 18pt

\def\ki{\Cal K}

\headline{\centerline{\sixrm 
Busch \&\ Lahti --- The Standard Model of Quantum Measurement Theory
\hfill \folio}}

\topinsert
\noindent{\sixrm Foundations of Physics 1996}
\endinsert

\title{\tit The Standard Model of Quantum Measurement Theory:}\\ 
{\tit History and Applications}
\endtitle

{\advance\baselineskip by -8pt
\centerline{{\gross Paul Busch}}
\centerline{\eightrm Department of Applied Mathematics}
\centerline{\eightrm The University of Hull}
\centerline{\eightrm Hull, HU6 7RX, England}
\centerline{\eightrm E-mail: p.busch\@ maths.hull.ac.uk}
\vskip 12pt
\centerline{\gross Pekka J.\ Lahti}
\centerline{\eightrm Department of Physics}
\centerline{\eightrm University of Turku}
\centerline{\eightrm SF-20500 Turku 50, Finland}
\centerline{\eightrm E-mail: pekka.lahti\@ utu.fi}

}

\vskip .4truecm

{\narrower
\advance\baselineskip by -4pt
\eightrm
\noindent
 The {\eightsl standard model} of the quantum theory of
measurement is based on an interaction Hamiltonian in which
the observable-to-be-measured is multiplied with some
observable of a probe system. This simple {\eightsl Ansatz}
has proved extremely fruitful in the development of the
foundations of quantum mechanics. While the ensuing type of
models has often been argued to be rather artificial, recent
advances in quantum optics have demonstrated their prinicpal
and practical feasibility. A brief historical review of the
{\eightsl standard model} together with an outline of its
virtues and limitations are presented as an illustration of
the mutual inspiration that has always taken place between
foundational and experimental research in quantum physics.

}

\vskip 3pt

\subheading{1. Introduction}

\noindent
 In Chapter Three of his classic {\it The Philosophy of
Quantum Mechanics}$^{(1)}$, Max Jammer reviews the debate
about the interpretation of the indeterminacy relation for
position and momentum. He points out `that of all the
interpretations listed ... the following two proved most
important for the development of quantum mechanics:

1.\ The {\it nonstatistical interpretation} I$_1$ according
to which it is impossible, in principle, to specify
precisely the simultaneous values of canonically conjugate
variables that describe the behavior of single (individual)
physical systems,
\vskip 0pt
2.\ The {\it statistical interpretation} I$_2$ according to
which the product of the standard deviations of two
canonically conjugate  variables has a lower bound given by
$h/4\pi$.'

 He then proceeds to show that under the
measurement-theoretic assumption of the measurements
involved being of the first kind, that is, repeatable, I$_1$
is a consequence of I$_2$. The Chapter concludes
with the observation `that certain modern developments have
cleared the way for the establishment of theories on the
simultaneous \mt\ of incompatible observables', but it is
pointed out that `little consensus has been reached on the
very basic issues of such theories, primarily, it seems,
because of diverging definitions of ``simultaneous
measurements.'' It is certainly too early to form a balanced
judgement on the legitimacy and prospects of such theories.'

Since the time when these lines were written (around 1974)
much has happened in theoretical as well as experimental
respects. Experimentally, there have been a variety of
illustrations of the scatter relation, for example, by means
of diffraction experiments involving beams of photons,
electrons, or neutrons. On the conceptual side, developments
that had started in the mid-1960s have led to a wider concept
of observables which allows the formulation of joint
observables for collections of noncommuting  quantities.  Quite
independently, measurement-theoretical models were developed
which were intuitively interpreted as simultaneous
position-momentum measurements. For a long time, both the
conceptual innovations and the concrete models remained
largely unnoticed, mainly, because their intimate
relationship was not made explicit. In fact a new and wider
mathematical representation of observables would have to be
justified in terms of -- at least model -- applications; and
the models of joint \mts\ did not allow for any reasonable
explanation in terms of the conventional quantum mechanical
formalism. Meanwhile the situation has changed completely in
that the models have been satisfactorily linked with the new
observable concept,
and search has begun for possible experimental realisations
of the corresponding `simultaneous', or joint measurements.

It was a certain type of \mt\ model that played an essential
role in these developments: in those models the
observable-to-be-measured enters the interaction Hamiltonian
as a factor, multiplied by an \ob\ of some probe \sy. In
view of its importance, we propose to name this sort of
model the {\it standard model} of quantum \mt\ \ty.
It is the purpose of this contribution to survey the main
features of the standard model  and to show how it leads, in
natural ways, to the representation of observables as
positive operator valued measures 
as well as to realisations of joint measurements of
noncommuting observables.


\subheading{2. A brief history of the standard model}

 Jammer's observation of I$_2$ implying I$_1$ shows that a
theory of simultaneous measurements of canonically
conjugate, or more general noncommuting pairs of observables
must be based on {\it non-repeatable} measurements. We
observe that logically this gives room for the consideration
of a {\it positive} reformulation of the nonstatistical
interpretation I$_1$ into

3.\ The {\it individual interpretation} I$_3$ according to
which it is {\it possible}, in principle, to specify {\it
unsharply} the simultaneous values of canonically conjugate variables that
describe the behavior of  single (individual) physical systems.

The twofold task is thus defined to (1) introduce appropriate notions of
joint observable and joint measurement for noncommuting
quantities in accordance with the probabilistic structure of
quantum mechanics, and
(2) demonstrate the principal realizability of these notions by
formulating concrete joint measurement models. In sections 3 and 4 we
shall employ the standard model to demonstrate how these goals can be
and indeed have been reached. Beforehand we shall recall some instances
in the development of quantum mechanics where this sort of model has
played a role. It will be seen that 
a proper understanding of the models in question could not
be obtained on the basis of the conventional quantum
formalism but was made possible only after the notion of
observables was extended appropriately.

Interestingly, the (to our knowledge) very first
mathematically rigorous and physically concrete model of
quantum \mt\ \ty\ leads to a non-repeatable \mt\ of an {\it
unsharp} position observable.  In fact on the last two pages
of his book {\it Mathematische Grundlagen der
Quantenmechanik}$^{(2)}$, von Neumann formulates a \mt\
scheme as a coupling between two particles -- object and
probe, effected by an interaction Hamiltonian of the form
$\frac h{2\pi i}q\frac\partial{\partial r}$ (in modern
notation: $Q\otimes P_{1}$, the product of the object
position with the probe momentum). This coupling is shown to
establish correlations between the object's position $Q$ and
the `pointer observable' $Q_{1}$, the position of the
probe. The ensuing process is intuitively interpreted as a
\mt\ of the object's position. Nevertheless it is evident
that the correlations in question are imperfect and it turns
out that the statistics of the \mt\ are not exactly given by
the usual quantum mechanical probability distributions for
the object's position. Therefore, the scheme presented by
von Neumann as an illustration of his \mt\ theory rather
possesses the potential of demonstrating a limitation of
the very quantum axiomatics put forward by himself:
as we shall see in the next section, the
model is adequately described as a \mt\ not of an observable in
von Neumann's sense but of what is nowadays
known as an {\it unsharp} position observable. 

Instead of drawing this conclusion, the standard model has
since been regarded by many authors as a fairly good
exemplification of an ideal and repeatable \mt\ of a single \ob.
This view is not entirely wrong: we shall see that in the
class of standard models there are realisations of repeatable
\mts. In other cases there is a nonzero \mt\ inaccuracy, and the
\mt\ is no longer repeatable. Yet this inaccuracy is determined
by the indeterminacy of the pointer observable in the
probe's initial state and can therefore be made arbitrarily
small by appropriate preparation of the probe. In this
spirit Aharonov and Bohm$^{(3)}$ propose a standard model
form of a momentum \mt\ as an unsharp \mt\ of the
kinetic energy of a particle in order to disprove a certain
form of energy-time uncertainty relation. In this example it
can be shown that in the limit of small inaccuracies a
remarkable nondisturbance feature arises: near-eigenstates
of the \me{}d \ob\ are transformed into near-eigenstates
with the same distribution.$^{(4)}$ 
It is this property of the standard model which gives it a
central status in the investigation of {\it quantum
non-demolition \mt} schemes. The nondisturbance
properties of such \mt{}s almost require them to be of the
standard model type.\footnote{\advance\baselineskip by -4pt
\eightrm For a review of the study of quantum non-demolition
\mt\ schemes and their utilisation in weak signal detection,
see, e.g.\  Ref.\ 5.}

Rather than concentrating on the small-inaccuracy limit of
the standard model, one might consider whether the
unsharpness implemented by the preparation of the probe is
of any positive use. Indeed if the probe is in a pure state,
the pointer unsharpness is due to the quantum mechanical
indeterminacy of the pointer values; and one could expect
that this is what is needed in order to devise simultaneous
\mts\ of noncommuting quantities in the spirit of
interpretation I$_3$.  Intuitively, it is only a small step
to consider combining two measurement schemes of the kind
introduced by von Neumann, one for a particle's position,
another one for its momentum, and to see if the resulting
process would serve as a joint measurement. More precisely,
one would couple two probes {\it simultaneously} with the
same particle by introducing an interaction Hamiltonian of
the form $\lambda Q\otimes P_1\otimes I_2+\mu P\otimes
I_1\otimes P_2$ ($\lambda,\mu$ being coupling constants).
This would amount to an extension of the standard model so
as to include the \mt\ of several \obs\ of one object.  Such
a model was first formulated by Arthurs and Kelly$^{(6)}$ 
in 1965.  It is interesting to note that a
variant of the Arthurs-Kelly model was studied independently
in  DeWitt's Varenna lecture of 1970$^{(7)}$, who also makes
extensive use of the standard model in illustrating the
many-worlds interpretation of quantum mechanics.  Still, at
that time the term {\it simultaneous \mt} could only be used
in an informal sense since no formal counterpart to the
measurement outcomes and their statistics was known that
could have served as a representation of the measured
observable in this model.  Accordingly, rather than
analysing the \mt\ statistics and drawing the conclusion
that they cannot be represented by spectral \me{}s of some
\sop{}s, these authors, as well as many others afterwards,
base their \mt\ criterion just on the reproduction of
expectation values and variances. This procedure has the
disadvantage that it does not single out a particular
observable as the \me{}d one. We shall see in section 4 that
the Arthurs-Kelly \mt\ scheme does determine a unique {\it
phase space observable} as the one whose statistics equals
the pointer statistics. This opens the way for a rigorous
\mt\ theoretical justification of Heisenberg's individual
interpretation I$_3$ of the indeterminacy relation.$^{(8)}$

The possibility of adjusting the measurement inaccuracy
within the family of standard models brings about a new
approach towards a new understanding of the classical domain
within quantum mechanics. For example, the Arthurs-Kelly
model of a phase space \mt, if appropriately generalized so
as to allow for arbitrary pointer preparations, is found to
display features typical of a classical \mt\ situation in
the limit of very large position and momentum
inaccuracies.\footnote{\eightrm For details, see Ref.\ 9.}
The importance of quantum \mts\ with large inaccuracies was
also recognized by Aharonov et al$^{(10)}$ when they
discovered a `surprising quantum effect' of classically
possible but quantum mechanically nonexistent readings in a
standard model treatment of a spin \mt\ with large pointer
variances. 

Turning now to the question of the experimental realization
of the standard model, it is tempting to refer to the
Stern-Gerlach experiment as perhaps the first example that
comes close to this type of measurement. However, the usual 
textbook description, which is based essentially on the
interaction Hamiltonian $H= \mu s_z\otimes(B_o+bz)$ and the
momentum $p_z$ as the pointer, can at best be
regarded as a caricature of the real situation as the
magnetic field is far more complicated than is assumed in
the above and indeed ought to satisfy Maxwell's 
equations.\footnote{\eightrm For an outline of a more
realistic account and some relevant references, see Ref.\ 9.}

The question whether the {\it phase space \mt} version of
the standard model allows for experimental realizations has
been answered in the positive only rather recently in the
context of quantum optics. Observing that the
phase and amplitude quadrature components of a photon field
are canonically conjugate quantities, Stenholm$^{(11)}$
showed that the Arthurs-Kelly coupling arises as an
approximation to a realistic quantum optical interaction.
Another realization that involves only one probe system is
obtained by a simple beam splitter coupling between a signal and
local mode, followed by a homodyne detection. This model was
derived by Leonhardt and Paul$^{(12)}$ as a simplification of
a more sophisticated phase \mt\ arrangement proposed by Noh,
Fougeres and Mandel.$^{(13)}$

\subheading{3. Quantum measurement theory and the standard
model}

\noindent
From the present-day point of view,
measurement theory as introduced by von Neumann was
restricted to the case of ideal measurements of sharp observables
represented by discrete self-adjoint operators.$^{(14)}$ 
We shall briefly recall this case in order to introduce  the ideas
of the measurement theory as employed in its standard model.

Let \s\ be the system on which a measurement is to be performed,
and assume that this system is represented by a (complex, separable)
Hilbert space $\hi$.
Assume further that one intends to measure an observable of \s\ given
by a discrete self-adjoint operator $A=\sum a_iP_i$, with the eigenvalues
$a_i$ and with the associated eigenprojections $P_i$, $i=1,2,\cdots$.
If \s\ is in a vector state $\fii$ (a unit vector of $\hi$), then
$p^A_\fii(a_i) = \ip{\fii}{P_i\fii}$
is the probability that a measurement of $A$ leads to the result
$a_i$.
Consider now another system \a, the measuring apparatus or the probe system, 
associated with
the Hilbert space $\ki$, and assume that it is prepared independently
in a vector state $\phi$, which is taken to be fixed. Let 
$U:\hi\otimes\ki\to\hi\otimes\ki$ be a unitary operator with the
property
$$
U(\fii\otimes\phi)   
= \sum P_i\fii\otimes\phi_i\tag 1
$$
where $\{\phi_i\}$ is a fixed set of mutually orthogonal
unit vectors in $\ki$.
Let $Z =\sum z_iZ_i$ be an observable of \a, the pointer observable,
such that $Z\phi_i = z_i\phi_i$ for all $i=1,2\cdots$. Let
$\pas(\pee\us)$ be  the partial trace of the vector state $\us$ over the
object Hilbert space $\hi$. 
We call it the (reduced) state of \a\ after the measurement.
Since $U$ and $\phi$ are assumed to be fixed,
we denote this state as $W(\fii)$.  Clearly, $W(\fii) = \sum p^A_\fii(a_i)
\pee{\phi_i}$, and one has for any $i$ and for all $\fii$,
$$
p^A_\fii(a_i) = p^Z_{W(\fii)}(z_i),\tag 2
$$
showing that the measurement outcome probabilities for $A$ in any
initial state $\fii$ of the object system  are recovered as the distribution 
of the pointer values  in the final apparatus state $W(\fii)$. In this sense
the system \a, with the Hilbert space $\ki$, its initial preparation $\phi$, 
the pointer observable $Z$, 
and the measurement coupling $U$ constitute a measurement of $A$.
We let $\Cal M = \vas\ki,\phi,Z,U\oik$ denote this measurement.
It is obvious that the items in this 4-tuple can be 
generalised and altered in many ways to produce various kinds
of measurements of $A$. The essential point is that any such $\Cal M$
is to fulfill the condition (2) in order serve as a measurement of 
the observable $A$.
The particular measurement scheme sketched above  is due to von
Neumann and it has a number of special properties. It is a first kind,
repeatable, and ideal measurement of  the sharp discrete observable $A$.

There is another reading of condition (2). 
 Given {\it any} measurement scheme
$\Cal M = \vas\ki,\phi,Z,U\oik$, Equation (2), when stipulated to
hold for all $\fii$ and for all $i$, 
defines the observable measured by this scheme. We shall go on to
apply this point of view  to the case of the standard measurement couplings
$$
U  \ =\ e^{i\lambda A\otimes B},\tag 3
$$
where $A$ is the observable intended to be measured,
$\lambda$ is a coupling constant and $B$ is an observable of \a.

Assume still that $A$ is discrete, $A = \sum a_iP_i$. 
In this case  $U$ may be written in the form
$U = \sum P_i\otimes e^{i\lambda a_iB}$ so that for any $\fii$
and for (fixed) $\phi$,
$$
\us
= \sum P_i\fii\otimes\phi_{\lambda a_i}^B,\tag 4
$$
where we have introduced the unit vectors
$\phi_{\lambda a_i}^B := e^{i\lambda a_iB}\phi$. 
Let the pointer observable be denoted by  $Z=\sum z_i Z_i$,
where the $z_i$ are the distinct eigenvalues, with the
mutually orthogonal eigenprojections $Z_i$ satisfying $\sum
Z_i=I$. We shall also use  notations like $Z:z_i\mapsto Z_i$
to indicated the associated positive operator valued (here:
spectral) measure.
Observe that the vectors $\phi_{\lambda a_i}^B$ need 
not be be mutually orthogonal, 
nor eigenvectors of $Z$. In any case, the condition
$$
p^E_\fii(a_i) :=  p^Z_{W(\fii)}(z_i),\tag 5
$$
when stipulated to hold for all $\fii$ and for all $i$,
defines the measured observable $E:a_i\to E_i$. 
Note that the relation (5) also induces a {\it pointer
function} $z_i\mapsto a_i$, so that a reading $z_i$ uniquely
indicates a value $a_i$ of the observable to be measured.
Since the final apparatus state has the form
 $W(\fii) = \sum_i p^A_\fii(a_i)\pee{\phi_{\lambda a_i}^B}$, 
a direct computation gives
$$
E_i = \sum_j p^Z_{\phi_{\lambda a_j}^B}( z_i)P_j.\tag 6
$$
The operators $E_i$ are positive and bounded by the unit
operator, $O\leq E_i\leq I$, and they sum up to the unit operator,
$\sum E_i = I$. 
This is to say that the mapping $E:a_i\mapsto E_i$
constitutes a (discrete) positive operator
valued measure. Furthermore, the structure of the operators $E_i$ show 
that they are weigthed means of the spectral projections of  $A$:
the measured observable $E$ is a smeared version of the sharp observable $A$.
The question at issue is whether one can choose $B$, $\phi$, and $Z$
such that $E_i = P_i$, so that the measured observable equals $A$.
Before we turn to  this question, let us observe that the measurement $\Cal M =
\vas\ki,\phi,Z,e^{i\lambda A\otimes B}\oik$
has the following two  properties:
the measured observable is {\it commutative}, that is,
$$
E_iE_j = E_jE_i\tag 7
$$ for all $i,j=1,2,\cdots$, and the measurement is
of the {\it first kind}, that is, the probability for a given outcome of the
measured observable is the same both before and after the measurement,
$$
p^E_\fii(a_i) = p^E_{T(\fii)}(a_i)
\text{ for all } \fii \text{ and all }i,\tag  8
$$
where $T(\fii) =\pa\big(\pee\us\big)$ is the
(reduced) state of \s\ after the measurement.

We demonstrate next that for $A=\sum a_iP_i$
one can in general choose $B$, $\phi$, and $Z$ such
that $E_i = P_i$ for all $i$.  To this end we need to ensure
that $Z_i\phi_{\lambda a_i}^B = \phi_{\lambda a_i}^B$ for
all $i$, which then entails automatically that
$\big\langle{\phi_{\lambda a_i}^B}|{\phi_{\lambda
a_j}^B}\big\rangle  = 0$, for $j\ne i$.
We take 
the probe system to be a particle moving in one-dimensional space, 
so that $\ki=L^2(\Rea)$,
and couple $A$ with its momentum $P_1$ via
$U=e^{-i\lambda A\otimes P_1}$.
Since the momentum generates translations of the position, it is
natural to consider choosing the position 
$Q_1$ conjugate to $P_1$ as the pointer observable. 
An initial state $\fii\otimes\phi$ of the object-probe system  is 
then transformed into $\sum_i P_i\fii\otimes \phi_{\lambda a_i}^{P_1}$. 
Using the position representation (for \a) one has 
$\phi_{\lambda a_i}^{P_1}(x)=
\phi(x-\lambda a_i)$. Assuming that the spacing between
the eigenvalues  $a_i$ of $A$ 
is greater than $\frac\delta\lambda$ and that $\phi$ is supported in 
$\bigl(-\frac\delta{2},\frac\delta{2}\bigr)$, then the 
``pointer states" $\phi_{\lambda a_i}^{P_1}$ are
supported in the mutually disjoint sets 
$\lambda I_i$, where
$I_i =\bigl(a_i-\frac\delta{2\lambda},a_i +\frac\delta{2\lambda}\bigr)$. 
This suggests to finally specify the pointer observable
$Z:z_i\mapsto Z_i$ to be a discretised position observable,
that is, $z_i:=\lambda a_i$ and $Z_i=E^{Q_1}(\lambda I_i)$, the spectral
projection of $Q_1$ associated with the interval $\lambda
I_i$. Then Equation (6) gives
$$
E_i \ = \
\sum_j  p^{Q_1}_{\phi_{\lambda a_j}^{P_1}}(\lambda I_i)\,P_j\ =\ P_i,
\tag 9
$$
 for each $i$.
It follows that the observable meas\-ured by this scheme is indeed $A$. 
Clearly, this measurement is just a realisation of the above 
abstract measurement scheme of von Neumann.

Let us still consider the coupling (3), but now without assuming that $A$ is
discrete. Using the spectral decomposition of $A$, $A =\int_{\bo R}
aP(da)$, one may still write $U$  in the form
$U\ =\ \int_\Rea\, P(da)\otimes e^{ia\lambda B}.$
An initial state $\fii\otimes\phi$ of the object-probe system transforms now
$$
\us \ =\ 
\int_\Rea \, P(da)\,\fii\otimes \phi_{\lambda a}^B,\tag 10
$$
with the notation $\phi_{\lambda a}^B := e^{ia\lambda B}\phi$,
and the final apparatus state gets the canonical form 
$$
W(\fii)\ =\ \int_\Rea p^A_\fii(da)\,\pee{\phi_{\lambda a}^B}.\tag 11
$$
If the pointer observable is a self-adjoint operator $Z$, then
the meas\-ured observable $E$ is obtained from the condition
$$
p^E_\fii(X) = p^Z_{W(\fii)}(\lambda X),\tag 12
$$
holding for all  (Borel) subsets  of the real line
 and for all initial vector states $\fii$ of \s. 
This gives 
$$
E(X) 
\ =\  \int_\Rea\, p^{Z}_{\phi_{\lambda a}^B}(\lambda X)\, P(da).
\tag 13
$$
Clearly, $O\leq E(X)\leq I$, and $E(\bo R) = I$, so that again
the mapping $E:X\mapsto E(X)$ is  a positive operator valued measure.

The structure of the operators $E(X)$  shows that  the actually
meas\-ured observable $E$ is not the observable $A$, 
but a smeared version of it. Again, we observe that the measured
observable $E$ is {\it commutative}, i.e., for all real Borel
sets $X,Y$, 
$$
E(X)E(Y) = E(Y)E(X), \tag 14
$$
and the measurement is of the {\it first kind},
$$
p^E_\fii(X) = p^E_{T(\fii)}(X),\ \text{ for all } \fii 
\text{ and for all sets } X,
\tag 15
$$
where $T(\fii)$ is the final state of \s.

The fact that for any $\phi$ and for all $Z$ the measurement
$\vas\ki,\phi,Z,e^{i\lambda A\otimes B}\oik$ is always of
the first kind has a remarkable implication: the measured
observable $E$ is never $A$ unless $A$ is discrete.  Indeed
it is known that a first kind measurement of a
sharp observable (that is, an observable represented by a
self-adjoint operator) is also repeatable and an observable
defined by a repeatable measurement is discrete.$^{(14)}$
Thus if $A$ is not discrete the measured observable is
always an unsharp version of $A$. 

We shall illustrate this important result by  taking $A$
to be (a Cartesian component of) the position of \s, $A =Q$. 
Choosing  $B=-P_1$ and $Z=Q_1$, we get for any $\phi$
$$
E(X) \ =\ \int_{\bo R} p^{Q_1}_{\phi_{\lambda q}^{P_1}}(\lambda X)\,
E^Q(dq), \tag 16
$$
where $E^Q$ denotes the spectral measure of $Q$, $Q=\int_{\bo R}qE^Q(dq)$.
Using the respective spectral representations,
the operators $E(X)$ assume the form
$$
E(X)\, =\,
 \iint\,
\left|\phi(q'-\lambda q)\right|^2 \,\chi_{{}_{\lambda X}}(q') dq'\,E^Q(dq)\, 
=\, \chi_{{}_X}*e(Q), 
\tag 17
$$
where the function  
$(\chi_{{}_X}*e)(y)=\int \chi_{{}_X}(x) e(y-x)dx$ 
is  the convolution of the characteristic function $\chi_{{}_X}$ with the
confidence function  
$e(x)\,:=\,\lambda\big|\phi(-\lambda x)\big|^2.$ 
Since $e$ cannot be a delta-function, the meas\-ured observable 
$$
E:X\mapsto \chi_{{}_X}*e(Q)\tag 18
$$
 is an unsharp position and not
the sharp  one $E^Q:X\mapsto \chi_{{}_X}(Q)$.
The standard ``position measurement" 
$\vas L^2(\bo R),\phi,Q_1,e^{-i\lambda Q\otimes P_1}\oik$
introduced by von Neumann thus 
determines always an unsharp position, where the ``unsharpness
parameter" $e$ depends on the preparation $\phi$ of the probe and of
the coupling constant $\lambda$.

The fact that the measured observable is an unsharp position 
and not the sharp one can be illustrated in terms of the
variance of the measured observable $E$, which is always greater than the 
variance of $Q$,
$$
\text{Var}\,(E,\fii) = \text{Var}\,(Q,\fii) + \frac 1{\lambda^2}
\text{Var}\,(Q_1,\phi).\tag 19
$$
The ``noise term" $\frac
1{\lambda^2}\text{Var}\,(Q_1,\phi)$, which
reflects the probe system's quantum nature,  
can be made small by an appropriate
choice of $\phi$, but it can never be eliminated. Only in the limit
of strong coupling $\lambda\to\infty$ would one have
$\text{Var}\,(E,\fii)\to\text{Var}\,(Q,\fii)$.

\subheading{4. A joint position-momentum
measurement model}

\def\la{\lambda}  \def\ot{\otimes}          \def\e0{E^{e_o}}
\def\f0{F^{f_o}}  \def\po{{\phi_1}}         \def\pt{{\phi_2}}
\def\xy{X\times Y}          \def\sqp{S_{qp}}

\noindent
 In this section we shall provide a model illustration of the
important fact that the \mt\ noise occurring in the von
Neumann model and the ensuing unsharp observables may play a
fundamental role rather than representing marginal
imperfections. To this end we shall combine a standard
unsharp position measurement
$\vas\ki_1,\phi_1,Q_1,\allowmathbreak e^{-i\lambda Q\otimes
P_1}\oik$ with a standard unsharp momentum measurement
$\vas\ki_2,\phi_2,P_2,e^{i\mu P\otimes Q_2}\oik$ to yield a
joint position-\-momentum measurement. Thus we consider a
meas\-uring apparatus consisting of two probe \sy{}s, $\Cal
A=\Cal A_1+\Cal A_2$, with initial state $\po\ot\pt$.  It
will be coupled to the object system \s, originally in state
$\fii$, by means of the interaction $$ U\ :=\
\exp\left(-\tst{\frac i\hbar}\,\la\, Q\ot P_1\ot I_2\,+\,
\tst{\frac i\hbar}\,\mu\, P\ot I_1\ot Q_2\right).\tag 20 $$
(In this section we let Planck's constant $\hbar$ explicitly
appear in the formulas.)
The coupling (20) changes the state of the object-apparatus system
$\Psi_o\equiv\fii\ot\po\ot\pt$
into $\Psi=U\Psi_o$ which in the position representation reads
$$
\Psi(q,\xi_1,\xi_2)\ =\  
\fii(q+\mu\xi_2)\,
\po(\xi_1-\la q-\tst{\frac{\la\mu}2}\xi_2)
\,\pt(\xi_2).\tag 21
$$

The meas\-ured observable $G$, which is a positive operator
valued measure on the (Borel) subsets of $\Rea\times\Rea$,
is determined from the probability reproducibility conditions
$$
\ip{\fii}{G(\xy)\,\fii}\ 
:=\ \left\langle \Psi\big|
I\ot E^{Q_1}(\la X)\ot E^{P_2}(\mu Y)\,\Psi\right\rangle,
\tag 22
$$
which are to hold for all initial object states $\fii$ and
for all outcome sets $X,Y$.
One obtains:
$$
G(\xy)\ =\ {{\frac 1{2\pi\hbar}}}
\int_{\xy}\sqp\,dq\,dp.
\tag 23
$$
The operators $\sqp$ are positive and have trace 1, and they
are phase space translates of an operator $S_0$ with the
same properties. $S_0$ depends on the initial probe states
as well as the coupling constants. As it is rather tedious to
write out this connection explicitly, we shall only detail
those aspects that are relevant to  the present argument.

The question at issue now is what reasons can be given for
interpreting the present \mt\ scheme as a joint \mt\ of
position and momentum. There are four arguments to be put
forward here. First, the basic idea formalized above was to
apply two \mt\ procedures simultaneously to the same object
system, and this should  be expected to yield a joint
\mt\ of the two quantities determined by the separate
schemes. Thus we have formulated what one would rightly call a
{\it simultaneous} \mt. 

Second, the \ob\ $G$ defined according to (22), (23) by the
simultaneous application of the position and momentum \mt\ schemes
should represent a
{\it joint observable} for position and momentum. And this
is true in the sense that the probability \me{}s given in
(22) are proper joint probabilities for a pair of unsharp
position and momentum observables. Equivalently, the
 marginals of the phase space \ob\ (23) are  unsharp
\pos\ and \mom\ \obs: 
$$
G(X\times \Rea)\, =\, E^e(X)\, =\, \chi_X*e(Q),
\quad G(\Rea\times Y)\, =\, F^f(Y)\, =\, \chi_Y*f(P).\tag 24
$$
It is straightforward to determine the explicit forms of the confidence
functions $e,f$:
$$\aligned
\quad e(q)\ &=\   \langle q|S_0|q\rangle\ =\ 
\int dq'\,\big|\phi^{(\la)}_1({\ts12}q'-q)\big|^2\,
 \big|\phi^{(\mu)}_2(q')\big|^2\ =\
 e_o*\big|\pt^{(\frac{\mu}{2})}\big|^2(q),\\
\qquad f(p)\ &=\   \langle p|S_0|p\rangle\ =\ 
\int dp'\,\big|\hat\phi^{(\mu)}_2({\ts12}p'-p)\big|^2\,
\big|\hat\phi^{(\la)}_1(p')\big|^2\  
=\ f_o*\big|\hat\phi^{(\frac{\la}2)}_1\big|^2(p).\\
\endaligned\tag 25
$$
Here we have introduced the  scaled functions
$$
\phi^{(\la)}_1(\xi_1)\ :=\ \sqrt\la\,\po(\la \xi_1),\qquad
\phi^{(\mu)}_2(\xi_2)\ :=\ 
{\frac 1{\sqrt\mu}}\,\po\left(\tst{\frac 1\mu}\xi_2\right).\tag 26
$$
 The functions $e_o$ and $f_o$ are the confidence functions
of the original single
\mt{}s which can be recovered from the present joint \mt\ model by
switching off one ($\mu=0$) or the other ($\la=0$) coupling.  
As indicated by the convolution structure, the original
undisturbed inaccuracies are each changed due to the presence of the other
device. In other words the simultaneous application of the meas\-uring
devices for $\e0$ and $\f0$ is not a joint \mt\ of this
original pair of unsharp position and momentum  but
rather of  coarse-grained versions $\fqf$ and $\fpg$ of them.

Third, the combination of position and momentum amounts to
considering {\it phase space} as the value set of an
observable. But the phase space is characterized as a
homogeneous space of the isochronous Galilei group.
Accordingly, a joint observable for position and momentum
should also be a {\it phase space observable}, that is, it
should have the proper covariance under space rotations,
space translations and Galilei boosts. In our
one-dimensional case we need not take into account the
rotation covariance. It turns out that the positive operator
valued measure $G$ of  Eq.\ (23) does have the required 
phase space translation covariance and thus qualifies as a
phase space \ob.

Finally, one may require that any \ob\ should allow for
approximately {\it nondisturbing} \mts\ that are capable of
registering the (approximate) values of the \ob, without
changing that value, whenever the system is in a
near-eigenstate. This intuitive idea and postulate can be
formalized and it is found that the present scheme does have
the corresponding nondisturbance property.$^{(9)}$

The fact that the confidence functions $e,f$ are position and momentum
distributions in one and the same "state" $S_0$ [Eq.\ (25)] immediately implies that
the \mt\ unsharpnesses represented by these functions satisfy an
uncertainty relation. Nevertheless, it is instructive to see the
dynamical mechanism at work in the
present model that forces this relation to arise due to the
mutual influence of the two \mt{}s being
carried out simultaneously. This becomes manifest in the variances of $e$ and
$f$:
$$
\aligned
\var(e)\, &=\, 
{\frac 1{\la^2}}\,\var(Q_1,\po)\,+\,
{\frac{\mu^2}4}\,\var(Q_2,\pt),\\
\var(f)\, &=\,
{\frac 1{\mu^2}}\,\var(P_2,\pt)\,\,+\,
{\frac{\la^2}4}\,\var(P_1,\po).\\
\endaligned\tag 27
$$
There are two ways to make the `undisturbed' variances (the first terms)
small: either by choosing large coupling constants or by preparing
`pointer' states having sharply peaked distributions $|\po|^2$, 
$|\hat\pt|^2$. Both options have the same consequence: they produce
large contributions to the other quantity's unsharpness (the second terms).
Thus there is no way of getting both quantities $\var(e)$ and $\var(f)$ 
small in one and the same experiment. 
Let us evaluate the product of the variances,
$$\aligned
\quad &\var(e)\cdot \var(f)\ = \ \Cal Q\,+\,\Cal D,\\
\quad &\Cal Q\ :=\ {\frac 14}\,\var(Q_1,\po)\,\var(P_1,\po)\,+\,
             {\frac 14}\,\var(Q_2,\pt)\,\var(P_2,\pt),\\
\quad &\Cal D\ :=\ {\frac 1{\,\,\la^2\mu^2}}\,\var(Q_1,\po)\,\var(P_2,\pt)\,+
\,{\frac{\,\,\la^2\mu^2}{16}}\,\var(Q_2,\pt)\,\var(P_1,\po).\\
\endaligned\tag 28
$$
Making use of the uncertainty relations $\var(Q_k,\phi_k)\,\var(P_k,\phi_k)\ge
\hbar^2/4$ for the two probe \sy{}s, we find that both terms $\Cal Q,\Cal
D$ can be estimated from below. Putting $x:=16\,\var(Q_1,\po)\,\var(P_2,\pt)/
(\la\mu\hbar)^2$, we obtain:
$$\aligned
\Cal Q\ &\ge\ \frac 14\left(\frac{\,\,\hbar^2}4+\frac{\,\,\hbar^2}4\right)\ =\ 
\frac{\,\hbar^2}8,\\
\Cal D\ &\ge\ \frac{\,\,\hbar^2}{16}\,\left(x+\frac 1x\right)\ \ \ \ge\ 
\frac{\,\,\hbar^2}8.\\
\endaligned\tag 29
$$
This shows finally that 
$$
\var(e)\cdot \var(f)\ = \ \Cal Q\,+\,\Cal D\ \ge
\frac{\,\,\hbar^2}8+\frac{\,\,\hbar^2}8\ =\ \frac{\,\,\hbar^2}4.\tag 30
$$
 It is remarkable that either one of the terms $\Cal Q$ and $\Cal D$
suffices to provide an absolute lower bound for the uncertainty
product. Hence there are two sources of inaccuracy that give rise to an
uncertainty relation.  Neglecting $\Cal D$, it would be simply the
uncertainty relations for the two parts of the apparatus 
which forbids making the term
$\Cal Q$ arbitrarily small. This is in the spirit of Bohr's argument
according to which it is the quantum nature of part of the
meas\-uring device that makes it impossible to escape the uncertainty
relation. Note that the two terms occurring in $\Cal Q$ each refer to one of
the probe \sy{}s, and they contribute independently to the lower bound for
$\Cal Q$; furthermore no coupling parameters appear in $\Cal Q$. 
There is no trace of a mutual influence between the two \mt{}s
being carried out simultaneously.
On the other hand neglecting the term $\Cal Q$, one would still
be left with the two contributions collected in $\Cal D$, the combination
of which has again a lower bound. The terms in $\Cal D$ are products of 
variances and coupling terms associated with the two probe \sy{}s, showing
that $\Cal D$ reflects the mutual disturbance of the two \mt{}s. This is in
accord with Heisenberg's illustrations of the uncertainty relation. For
example if a particle is \me{}d so as to have a rather well-defined \mom,
then a subsequent \mt\ of position by means of a slit influences the effect
of the preceding \mom\ \mt\ to the extent required by the uncertainty
relation. 

Finally we should like to emphasise that the nature of the \mt\
`inaccuracy', or unsharpness, is determined by the preparations of the
apparatus. Insofar as the pointer \obs\ are indeterminate and not
merely subjectively unknown, this interpretation applies to the \mt\
uncertainties as well: each individual \mt\ outcome is intrinsically
unsharp, reflecting thereby a genuine quantum noise inherent in the \mt\
process, so that the inequality (30) should be properly called an
indeterminacy relation.

 One would expect that being able to perform phase space \mt{}s it should
also be possible to observe trajectories of microscopic particles. Thus one
may hope to achieve a detailed quantum mechanical account of the
formation of cloud or bubble chamber tracks. We shall indicate here that within
the present model  necessary conditions for such quasi-classical \mt\
behaviour are  ``macroscopically'' large
inaccuracies contributed by the device and, relative to the scale of
these inaccuracies, good localisation of the object.

 A {\it (quasi-)classical \mt\ situation} is characterised among others by the
possibility of observing a particle without necessarily influencing it. The
above phase space \mt\ model allows one to formalise this and some further
classicality conditions and to demonstrate their realizability. We shall
formulate four such requirements.

First, it should be admissible to think of the particle having
``arbitrarily
sharp'' values of \pos\ and \mom. This cannot be meant in an absolute sense
but only relative to the scale defined by the resolution of the means of
\mt.

\noindent{(C1)} {\it Near value determinateness.} 
$$
\var(Q,\fii)\ \ll\ \var(e),\qquad \var(P,\fii)\ \ll\ \var(f).\tag 31
$$
 Such states may be viewed as ``localised'' in phase space.

Next, the \pos\ and \mom\  \mt{}s should not disturb each other when
performed  jointly. This can be controlled by the variances (27) in terms
of the condition that the additional noise terms should remain negligible:

\noindent{(C2)} {\it Small mutual disturbance.} $\var(e)\simeq \var(e_o)$ and
$\var(f)\simeq \var(f_o)$; therefore
$$\aligned
{\frac 1{\la^2}}\,\var(Q_1,\phi_1)\ \gg\
{\frac{\mu^2}4}\,\var(Q_2,\phi_2),\\
{\frac 1{\mu^2}}\,\var(P_2,\phi_2)\ \gg\
{\frac{\la^2}4}\,\var(P_1,\phi_1).
\endaligned
\tag 32
$$

 Third, in view of the uncertainty relation (2.18) for the \mt\
inaccuracies it should be kept in mind that in a classical \mt\ the
imprecisions seem to be so large that no indication of Planck's constant 
can ever be observed.

\noindent{(C3)} {\it No limit of accuracy.} 
The \pos\ and \mom\ \mt\ inaccuracies can be made arbitrarily small:
$$
\var(e)\,\var(f)\ \gg\ \frac{\,\,\,\hbar^2}4.\tag 33
$$

Finally, since the properties to be \me{}d are practically determinate
in localized states,
one should expect that a \mt\ will not necessarily disturb the \sy\ but
will merely register the corresponding values. This is to say that the
\mt\ should be  approximately nondisturbing.

\noindent{(C4)} {\it Approximate nondisturbance.} Localised states [cf.\ (C1)]
should not be disturbed much in a joint \pos-\mom\ \mt.

These features are not mutually independent. It is evident
that (C1) implies (C3). The second property, (C2), gives somewhat more:
$$
\var(e)\,\var(f)\ \ge\ \var(e_o)\,\var(f_o)\ \gg\ \frac{\,\,\,\hbar^2}4.
\tag 34
$$ 
This shows that mutual
nondisturbance can only be achieved if from the outset one starts with
highly unsharp \mt{}s. If the requirements expressed in (C1)
and (C2) are somewhat strengthened,
one can prove by means of the present model that the approximating conditions
(C1) and (C2) are self-consistent, can be realised, and do indeed lead to 
(C4).$^{(9)}$

In the classical \mt\ situation described here, one is facing two kinds of
uncertainty which have to be interpreted quite differently. If one starts
with a localised though otherwise unknown state, then it is a matter of
{\it subjective ignorance} what the ``true values''
$(q_o,p_o)=\bigl(\langle Q\rangle_\fii,\langle P\rangle_\fii\bigr)$ 
of \pos\ and \mom\ are. The \mt\ will give some (point-like) outcome
$(q,p)$ most likely in a region around $(q_o,p_o)$, for which the probabilities
are non-negligible provided that $|q-q_o|\le n\sqrt {\var(e)}$,
$|p-p_o|\le n\sqrt{\var(f)}$, with $n$ of the order of unity. {\it Which}
result will come out is objectively undecided as the unsharpnesses
originate from the pointer indeterminacies. Hence with respect to the state
inference problem one is dealing with subjective uncertainties, while 
predictions of future \mt\ outcomes are objectively indeterminate.

These considerations reveal the decisive role of Planck's
constant $\hbar$ for the classical limit of \qm\ in a new sense. Only with
respect to meas\-uring instruments yielding macroscopic inaccuracies, Eq.\
(34), is it possible to neglect the \qu\ \mec{}al restrictions and to make
approximate use of the classical physical language as laid down in (C1--4).
It is remarkable that the indeterminacy product
$\var(Q,\fii)\var(P,\fii)$ of the object and
the products $\var(Q_k,\phi_k)\var(P_k,\phi_k)$ of 
the probe \sy{}s need not at all be small for quasi-classical \mt{}s. Thus
one can conceive of meas\-uring classical trajectories for microscopic
particles.

A more detailed analysis of the state changes incurred by the object
system shows that the present model  offers a continuous
transition between the two
extremes of nearly repeatable and nearly nondisturbing (first kind)
\mt{}s; thus it becomes evident
that these two ideals fall apart into mutually exclusive
options for unsharp \obs, while in the case of sharp \obs\ the two
concepts coincide.

\subheading{5.\ Conclusion: some reflections}

\noindent
 In the presentation of the standard model we have freely
used the two possible alternative readings of the {\it
probability reproducibility condition} (5): on one hand,
this relation serves as a criterion for a measurement scheme
to be a measurement of a given observable; on the other
hand, the observable actually measured by the scheme is
uniquely determined by (5). Thus, starting with a quantum
mechanical modelling of a measurement as a physical process,
one is inevitably led to introduce the measured observable
as a positive operator valued measure. From this point of
view, it is obvious that quite some experimental ingeniuity
is required for realising a \mt\ of a sharp observable.

It is remarkable that for a long time the quantum
measurement theory and the theory of positive operator
valued measures developed quite independently and largely
without taking notice of each other. That situation may be
characterized by saying that the modelling of measurements
would be doomed to be blind unless it was carried out in the
light of the concept of observables as positive operator
valued measures; and that the theory of general observables
would remain void of physical content unless it became
supplemented with concrete measurement models. Fortunately
the two have since been brought together into a fruitful
interaction. 

There are several possible reasons why the
standard model did not by itself give rise to the
generalization of the notion of observable required for an 
understanding of what was being measured. For example, von Neumann
came close to the conclusion in question when he calculated
the joint probability for the pointer and the observable to
be measured; but he only used this result to indicate that
the correlation between these quantities could be made
arbitrarily, though not absolutely, strong. This shows that
he identified the term {\it \mt} with {\it repeatable \mt},
since he concluded that due to the rather good correlations,
his model could be regarded as an approximate realization of
a \mt\ of position. In this view the intrinsic inaccuracy of
the \mt\ is regarded as an unwanted imperfection and not
seen as the key to a proper notion of a joint unsharp \mt.
Accordingly, von Neumann sketches a concept of joint \mt\
for position $Q$ and momentum $P$ where two commuting sharp
observables $Q',P'$ are constructed that are close to $Q$
and $P$, respectively, in some suitable topology (Ref.\ 2,
Section V.4). He regards this procedure as `purely
mathematical'. By contrast, the notion of phase space \mt\
reviewed in Section 4 is physically quite appealing and not
far from experimental realizability. 

Another reason why the standard model played no role in
introducing the general concept of observables may have been
the following.  In most of the papers dealing with variants
of the standard model, the Heisenberg picture was used
instead of the Schr\"odinger picture, which then gave rise
only to the conclusion that the pointer observable has the
same expectation value as the observable intended to be
measured. In addition the proportionality between the
Heisenberg operators representing the pointer observable and
the observable intended to be measured seemed to suggest
taking the latter for the measured \ob. However, as we
pointed out above, it is (only) the probability reproducibility
condition (5) that gives a thorough characterization of the
observable actually  measured; and this differs, in general,
from the intended \ob. To be fair, it was recognized that
the \mt\ schemes in question provided only an approximate
\mt\ of the object observable under consideration. Equation
(6) illustrates the precise relation between the \me{}d \ob\
and the underlying sharp \ob: the former is a smeared
version of the latter.

At this point it might appear as if positive operator valued
measures, as they emerge in standard model-type \mt{}s, are
not much more than a convenient description of imperfect, or
inaccurate \mt{}s of ``ordinary'', sharp observables. It is
true that this is one purpose of certain unsharp \obs. But
it must be noted that not all unsharp observables are
commutative, and when they are not, there are no underlying
sharp observables of which they were smeared versions. The
phase space observable of Section 4 is an example of a kind
of experimental question that does not allow for a
theoretical formulation on the basis of sharp observables
only. In this example it is crucial to realize that there is
a genuinely quantum mechanical source of unsharpness,
namely, the indeterminacy of the pointer observables in the
initial probe states. According to equations (19) and (28),
this can be interpreted in accordance with I$_3$ as quantum
noise inherent in the \mt\ outcomes so that the \mt\
uncertainties satisfy the Heisenberg indeterminacy relation.

The unsharpness inherent in observables that are not
represented as \sop{}s brings about yet another important
innovation into quantum physics.  The von Neumann model was
found to be an unsharp, non-repeatable \mt\ of position
(Section 3). Considering that the lack of repeatability is
related to the fact that the \me{}d \ob\ is continuous, one
might try to restore repeatability in this model by
discretising the pointer observable, as this would amount to
defining a discrete, though still unsharp, position
$X_i\mapsto E(X_i) = \chi_{_{X_i}}*e(Q)$. It turns out that
the measurement, while always of the first kind, still will
not be repeatable. The fact that for measurements of unsharp
observables the first kind-property is essentially weaker
than the repeatability property comes as an advantage in
some respects. Indeed repeatable measurements will produce
strong ``disturbances" in that they turn non-eigenstates
into eigenstates of the measured observable. By contrast,
the first kind-property is a ``nondisturbance" feature of a
measurement as it only ensures that the states are changed
so gently that the distribution of the values of the
measured observable does not change. It follows that one
gains more control over the magnitude of state changes for a
larger class of states than just eigenstates (if there are
any).   This nondisturbance property is
what one would expect to be present in a classical
measurement situation (Section 4): macroscopic, classical
\obs\ have fairly (though perhaps not absolutely)
well-defined values which can be detected without being
changed.

To conclude, we observe that the assessment of the role of
the \mt\ unsharpness in the standard model has shifted from
``marginal'', ``negligible'' or perhaps ``undesirable'' towards
``interesting'' or even ``crucial'' in some circumstances; and this
shift went along with the conception of joint measurements of
noncommuting quantities and of \mts\ that are less invasive
-- and yet sometimes more informative -- than sharp \mts.

\subheading{References}

{\advance\baselineskip by -4pt
\eightrm

\item{1.\ \ } M.\ Jammer. {\eightsl The Philosophy of Quantum Mechanics}
(Wiley, New York, 1974).

\item{2.\ \ } J.\ von Neumann. {\eightsl Mathematische Grundlagen der
Quantenmechanik} (Springer, Berlin, 1932); English transl.: {\eightsl
Mathematical Foundations of Quantum Mechanics} (Princeton University
Press, Princeton, 1955).

\item{3.\ \ } Y.\ Aharonov, D.\ Bohm, 
{\eightsl Phys.\ Rev.\ 122}, 1649 (1961).

\item{4.\ \ } P.\ Busch, {\eightsl Found.\ Phys.\ 20}, 33 (1990).

\item{5.\ \ } C.M.\ Caves, K.S.\ Thorne, R.W.P.\ Drever, V.P.\ Sandberg,
M.\ Zimmerman, {\eightsl Rev.\ Mod.\ Phys.\ 52}, 341 (1980).
D.F.\ Walls, in {\eightsl Symposium on the Foundations of Modern Physics
1993}, eds.\ P.\ Busch, P.\ Lahti, P.\ Mittelstaedt (World Scientific,
Singapore, 1993).

\item{6.\ \ } E.\ Arthurs, J.L.\ Kelly,
{\eightsl Bell Syst.\ Tech.\ J.\ 44},
725 (1965).

\item{7.\ \ } B.S.\ DeWitt, in {\eightsl
Foundations of Quantum Mechanics}, ed.\ B.\ d'Espagnat
(Academic Press, New York, 1972).

\item{8.\ \ } P.\ Busch, Doctoral Dissertation, Cologne
1982; English transl.: {\eightsl Int.\ J.\ Theor.\ Phys.\ 24},
63 (1985). P.\ Busch, P.J.\ Lahti, {\eightsl Philosophy of Science 52},
64 (1985).

\item{9.\ \ } P.\ Busch, M.\ Grabowski, P.\ Lahti, {\eightsl
Operational Quantum Physics} (Springer-Verlag, Berlin, 1995).

\item{10.\ } Y.\ Aharonov, D.Z.\ Albert, L.\ Vaidman, 
{\eightsl Phys.\ Lett.\ 124A}, 199 (1987); {\eightsl
Phys.\ Rev.\ Lett.\ 60}, 1351 (1988).

\item{11.\ } S.\ Stenholm, {\eightsl Ann.\ Phys.\
(N.Y.) 218}, 233 (1992).

\item{12.\ } U.\ Leonhardt, H.\ Paul, {\eightsl Phys.\
Rev.\ A47}, R2460 (1993); U.\ Leonhardt, {\eightsl
Phys.\ Rev.\ A48}, 3265 (1993).

\item{13.\ } J.W.\ Noh, A.\ Fougeres, L.\ Mandel, 
{\eightsl  Phys.\ Rev.\ A45}, 424 (1992); {\eightsl Phys.\ Rev.\ A46},
2840-2852 (1992).

\item{14.\ } P.\ Busch, P.\ Lahti, P.\ Mittelstaedt, {\eightsl
The Quantum Theory of Measurement} (Springer-Verlag, Berlin, 1991;
2nd edition in preparation).


\subheading{Footnotes}
   
\item{1.\ \ } For a review of the study of quantum non-demolition
\mt\ schemes and their utilisation in weak signal detection,
see, e.g.\  Ref.\ 5.

\item{2.\ \ } For details, see Ref.\ 9.

\item{3.\ \ } For an outline of a more realistic account and some
relevant references, see Ref.\ 9.

}
\bye